\documentclass[prc,notitlepage,letterpaper,showpacs]{revtex4-1}
\usepackage{color,graphicx}
\usepackage{soul}

\begin{document}

\title{Occupation probabilities from quadrupole moments in the Sn region}
\author{N. B. de Takacsy }
\affiliation{McGill University, Department of Physics, Montreal, Qc, Canada, H3A 2T8}

\date{26 February 2014}

\begin{abstract}
It is shown that a simple BCS model with a quadrupole-quadrupole interaction provides a consistent description of the measured quadrupole moments
 of a sequence of odd mass Sn and Cd isotopes and allows the extraction of the neutron single particle occupation probabilities.
\end{abstract}


\maketitle

\section{Introduction.}

In a recent high accuracy experimental study of electric quadrupole and magnetic dipole moments of a sequence of odd-mass Cd isotopes \cite{yordanov2013}
 it was noted that the electric quadrupole moments of the $\frac{11}{2}^-$ states in the Cd isotopes vary linearly with neutron number and 
are very simply and very well described in terms of a $h_{11/2}$ orbital sequentially occupied in the seniority scheme with a large neutron effective charge $e_n = 2.5$.
If this is so, the same simple physics should be manifest in the other presumed single quasi-particle states in Cd and, a-fortiori, in the the Sn isotopes which have
 a closed proton shell. The idea that the single neutron occupation probabilities can be so simply extracted from electromagnetic properties is sufficiently attractive
 to warrant a more detailed second look. The aim of this paper is to show that this can be done using a minimalist theoretical framework. 
Ultimately of course, large scale theoretical calculations, shell model or beyond-mean-field model,  are required to describe the structure of these nuclei,
 including their electromagnetic properties. The aim of this paper is much more modest, but the picture is nevertheless useful for its simplicty.

\section{The model.}

As theoretical framework, I use the spherical BCS model with a weak-coupling approximation to include quadrupole collectivity. 
In this model, \cite{ring-schuck-1980} the states in the odd-even isotopes with dominantly single quasi-particle parentage can be written as:

\begin{equation}
\mid j,m \rangle = \mid (1qp) jm \rangle + C_2 \mid j \otimes 2^{+};jm \rangle    \label{qp_state}
\end{equation}
where 
\begin{equation}
\mid (1qp) jm \rangle = \alpha^{\dagger}_{n \ell jm} \mid 0 \rangle    
\end{equation}
and 
\begin{equation} 
\mid j \otimes 2^{+};jm \rangle = \sum_{m',M'}   (j,m',2,M'\mid j,m) \alpha^{\dagger}_{n \ell jm'} \mid 2^{+},M' \rangle  
\end{equation} 

with the usual relation between the quasi-particle operators $\alpha^{\dagger}$ and the particle operators $a^{\dagger}$

\begin{equation} 
\alpha^{\dagger}_{n \ell jm} = u_{n \ell j} a^{\dagger}_{n \ell jm} - (-)^{j-m} v_{n \ell j} a_{n \ell j-m}
\end{equation}

The BCS ground state is $\mid0\rangle$, and $\mid2^{+}\rangle$ represents the first (collective) $2^+$ state which can usually be interpreted as a coherent superposition of two quasi-particle states. 
The coefficient $C_2$ is given in first-order perturbation theory by

\begin{equation}
C_2 = - \langle (1qp)jm \mid {\hat V}  \mid j \otimes 2^{+};jm \rangle / (E_2 - E_0) 
\end{equation}

I take the interaction potential ${\hat V}$ to be a quadrupole-quadrupole force \cite{bes-sorensen-1995}

\begin{equation}
{\hat V} = -\frac{1}{2} \chi_{pp}\sum_{\mu} (-)^\mu q_\mu(p) q_{-\mu}(p) - \frac{1}{2} \chi_{nn}\sum_\mu (-)^\mu q_\mu(n) q_{-\mu}(n)
           - \chi_{pn} \sum_\mu (-)^\mu q_\mu(p) q_{-\mu}(n)
\end{equation}

where p,n refer to protons and neutrons respectively and the one-body quadrupole operator is 

\begin{equation}
q_\mu = \sum_{n \ell jm} \left( \int \psi^{*}_{n\ell jm}  r^2 Y_{2,\mu}(\theta,\phi) \psi_{n\ell jm} d^3 {\vec r} \right) :a^{\dagger}_{n\ell jm} a_{n\ell jm}:
\end{equation}

For a neutron quasi-particle,  the expression for $C_2$ then becomes

\begin{equation}
C_2 = \frac{1}{E_2-E_0} \frac{1}{\sqrt{5(2j+1)}} (u^2_{n\ell j} - v^2_{n\ell j} ) \langle \psi_{n\ell j} \mid\mid r^2Y_2 \mid\mid \psi_{n\ell j} \rangle \langle 2^{+} \mid\mid \chi_{nn} q(n) + \chi_{pn} q(p) \mid\mid 0 \rangle 
\end{equation}

\section{The quadrupole moments}

To first order in $C_2$, the electric quadrupole moment of the weak coupling state in equation~\ref{qp_state} is
\begin{eqnarray}
Q_{chg}(j)  &=& \langle j,j \mid e_n q_0(n) +e_p q_0(p) \mid j,j \rangle  \\
            &=& \sqrt{\frac{16\pi}{5}}(2,0,j,j \mid j,j) \left[ (u^2_j - v^2_j)\frac{1}{\sqrt{2j+1}} \langle \psi_j \mid\mid e_n r^2 Y_2 \mid\mid \psi_j \rangle +
                 2C_2 \frac{1}{\sqrt{5}} \langle 2^{+} \mid\mid e_n q(n) + e_p q(p) \mid\mid 0 \rangle   \right] \nonumber                                           \\
            &=& \sqrt{\frac{16\pi}{5}} (2,0,j,j \mid j,j)(u^2_j - v^2_j)\frac{1}{\sqrt{2j+1}} \langle \psi_j \mid\mid r^2 Y_2 \mid\mid \psi_j \rangle    \nonumber  \\     
            & & \left[ e_n + \frac{2}{5} \frac{1}{E_2-E_0} \langle 2^{+} \mid\mid e_n q(n) + e_p q(p) \mid\mid 0 \rangle \langle 2^{+} \mid\mid \chi_{nn} q(n) + 
                \chi_{pn} q(p) \mid\mid  0 \rangle \right]        
\end{eqnarray}
Here $e_n$ and $e_p$ are the neutron and proton effective charges. In shell model calculations, their values are roughly $e_n \approx 0.5 e$ and $e_p \approx 1.5 e
$ \cite{brown2001}\cite{honma2004}, attributed mainly to the giant quadrupole resonance that is not explicitly included in the usual shell model. 
There are different ways of estimating the  strength of the effective quadrupole-quadrupole force, however $|\chi_{nn}|$ is always substantially smaller than $|\chi_{pn}|$ .
 Since a ratio $\chi_{nn} / \chi_{pn} \approx 0.3 $ is obtained from Hartree-Fock calculations with a Skyrme interaction \cite{terasaki2002},\cite{dobaczewski1988}, I feel justified
 in making the approximation that 

\begin{equation} 
  \langle 2^{+} \mid\mid e_n q(n) + e_p q(p) \mid\mid 0 \rangle  \propto \langle 2^{+} \mid\mid \chi_{nn} q(n) + 
                \chi_{pn} q(p) \mid\mid  0 \rangle
\end{equation}

For a neutron quasi-particle state, the quadrupole moment can then be written as 
\begin{eqnarray}
Q_{chg}(j)                                         
            &=& \sqrt{\frac{16\pi}{5}} (2,0,j,j \mid j,j)\frac{1}{\sqrt{2j+1}} \langle \psi_j \mid\mid r^2 Y_2 \mid\mid \psi_j \rangle (u^2_j - v^2_j) 
                \left( e_n + \frac{2}{E_2 - E_0} \left( \frac{{\bar \chi}}{e_n}\right)  B(E2;0^{+} \rightarrow 2^{+}) \right)   \nonumber        \\
            &=& Q_{sm}(j) (u^2_j - v^2_j) \left( e_n + \frac{2}{E_2 - E_0} \left( \frac{{\bar \chi}}{e_n}\right)  B(E2;0^{+} \rightarrow 2^{+}) \right) \label{q_moment} \\
            &=& Q_{sm}(j) (u^2_j - v^2_j) e_{n,tot}      
\end{eqnarray}
where the last equation defines the total effective charge $e_{n,tot}$ and the shell model quadrupole moment is 
\begin{eqnarray}
Q_{sm} &=& -\frac{1}{2} \frac{(2j-1)}{(j+1)} I^{(2)}_{n\ell j} \\
I^{(2)}_{n\ell j} &=& \int_0^\infty \mid R_{n \ell j} \mid^2  r^4 dr 
\end{eqnarray}
Note that the spectroscopic quadrupole moment normally reported in experiments is $ Q_s = Q_{chg}/e $. Provided that the parameter ${\bar \chi}$ can be estimated,
 equ.~\ref{q_moment} can be used to simply extract the neutron occupation probabilities from experimental data on quadrupole moments and E2 transition probabilities.

\section{The S\lowercase{n} and C\lowercase{n} isotopes }

The spectroscopic quadrupole moments and $B(E2)$ values for the Sn and Cd isotopes have been measured over a range of masses between the magic neutron numbers $N=50$ and $N=82$. 
Table 1 summarizes the experimental excitation energy and $B(E2 \uparrow)$ value  of the $2+$ states in the even-even isotopes  and Table 2 lists the measured spectroscopic 
quadrupole moments of the of the odd-neutron isotopes. The radial integrals $I^{(2)}_{n\ell j}$ are calculated using the shell model potential of reference~\cite{lojewski2001}. 
The strength of the quadrupole-quadrupole interaction is taken to be constant ${\bar \chi} = 0.3\times 10^{-3} MeV fm^{-4}$. This is chosen to yield reasonable 
values  of $u^2-v^2$ for the $11/2-$ states in the lightest Sn and Cd isotopes where the $0h_{11/2}$ is expected to be mostly empty and near $N=81$ where it should be filled. 
The factor $B(E2)/(E_2 - E_0)$ is taken to be the average of the experimental values in the even neutron isotopes above and below the odd neutron isotope being studied. 
When both are not available, then the nearest $B(E2)$ value and the average energy denominator is used.  The extracted occupation functions $u^2-v^2$ are given in Table 2 and
 shown in Figs.1 and 2. The uncertainty in $u^2-v^2$ is estimated from the uncertainties in the experimental $B(E2)$ and $Q_s$ and does not include the uncertainty in
 ${\bar \chi}$, nor the uncertainty in $e_n$.

\begin{table}[h]
\centering
\begin{tabular}{|l|l|l|r|l|l|l|l|r|l|}
\hline
Z & A & $E(2^{+})$ & $B(E2;0^{+}\rightarrow 2^{+})$ & Reference & Z & A & $E(2^{+})$ & $B(E2;0^{+}\rightarrow 2^{+})$ & Reference \\
  &   &   [MeV]  &   $[e^2 b^2]$                  &           &   &   &   [MeV]  &   $[e^2 b^2]$                  &           \\
\hline
48 & 98 & 1.395  &            &                     &   50 & 100 & $\approx 3$ &            &  \\
\hline
48 & 100 & 1.004 &            &                     &   50 & 102 & 1.472     &            &  \\
\hline
48 & 102 & 0.777 & 0.281 (45) & \cite{boelaert2007} &   50 & 104 & 1.26      & 0.163 (26) & \cite{doornenbal2013} \\
\hline
48 & 104 & 0.658 & 0.390 (14) & \cite{boelaert2007} &   50 & 106 & 1.207     & 0.209 (32) & \cite{kumar2010} \\
\hline
48 & 106 & 0.633 & 0.410 (20) & \cite{raman2001}    &   50 & 108 & 1.206     & 0.224 (16) & \cite{kumar2010} \\
\hline
48 & 108 & 0.633 & 0.430 (20) & \cite{raman2001}    &   50 & 110 & 1.212     & 0.226 (18) & \cite{kumar2010} \\
\hline
48 & 110 & 0.658 & 0.450 (20) & \cite{raman2001}    &   50 & 112 & 1.257     & 0.242 (8)  & \cite{kumar2010} \\
\hline
48 & 112 & 0.617 & 0.510 (20) & \cite{raman2001}    &   50 & 114 & 1.3       & 0.232 (8)  & \cite{kumar2010} \\
\hline
48 & 114 & 0.558 & 0.545 (20) & \cite{raman2001}    &   50 & 116 & 1.294     & 0.209 (6)  & \cite{kumar2010} \\
\hline
48 & 116 & 0.514 & 0.560 (20) & \cite{raman2001}    &   50 & 118 & 1.23      & 0.209 (8)  & \cite{kumar2010} \\
\hline
48 & 118 & 0.488 & 0.568 (44) & \cite{raman2001}    &   50 & 120 & 1.171     & 0.202 (4)  & \cite{kumar2010} \\
\hline
48 & 120 & 0.506 & 0.48  (6)  & \cite{raman2001}    &   50 & 122 & 1.141     & 0.192 (4)  & \cite{kumar2010} \\
\hline
48 & 122 & 0.569 & 0.41  (3)  & \cite{ilieva2013}   &   50 & 124 & 1.132     & 0.162 (6)  & \cite{allmond2011} \\
\hline
48 & 124 & 0.613 & 0.35  (4)  & \cite{ilieva2013}   &   50 & 126 & 1.141     & 0.127 (8)  & \cite{allmond2011} \\    
\hline
48 & 126 & 0.652 & 0.22  (2)  & \cite{ilieva2013}   &   50 & 128 & 1.169     & 0.080 (5)  & \cite{allmond2011} \\
\hline
48 & 128 & 0.645 & 0.16  (2)  & \cite{boning2012}   &   50 & 130 & 1.221     & 0.023 (5)  & \cite{kumar2010} \\
\hline
48 & 130 & 1.325 &            &                     &   50 & 132 & 4.041     & 0.11  (3)  & \cite{radford2005},\cite{beene2004}   \\
\hline
\end{tabular}
\caption{\label{tab:1/tc} Experimental $B(E2;0^{+}\rightarrow 2^{+})$ of a sequence of Cd and Sn isotopes.}
\end{table}

\begin{table}[h]                   
\centering                   
\begin{tabular}{|l|l|l|l|l|l|l|l|l|}                   
\hline                   
Z  & A   & E*=     & $J^{\pi}$ & $n\ell_{j}$ & $I^{(2)}_{n\ell j}$ & $Q_{s}(expt)$ & $(u^2-v^2)$ & Reference(expt) \\
\hline                   
   &     & $[MeV]$ &           &             & $[fm^2]$            & $[b]$         &             &                 \\  
\hline                   
48 & 103 & 0       & 5/2+  & $1d_{5/2}$  & 24.112 & -0.8(7)   & 1.30(114) & \cite{stone2011} \\
\hline                   
48 & 105 & 0       & 5/2+  & $1d_{5/2}$  & 24.365 & 0.43(4)   & -0.54(5)  & \cite{stone2011} \\
\hline                   
48 & 107 & 0       & 5/2+  & $1d_{5/2}$  & 24.614 & 0.601(3)  & -0.71(3)  & \cite{yordanov2013} \\
\hline                   
48 & 107 & 0.846   & 11/2- & $0h_{11/2}$ & 28.143 & -0.94(10) & 0.72(8)   & \cite{stone2011} \\
\hline                   
48 & 109 & 0       & 5/2+  & $1d_{5/2}$  & 24.864 & 0.604(1)  & -0.69(3)  & \cite{yordanov2013} \\
\hline                   
48 & 109 & 0.464   & 11/2- & $0h_{11/2}$ & 28.454 & -0.92(9)  & 0.68(7)   & \cite{stone2011} \\
\hline                   
48 & 111 & 0.245   & 5/2+  & $1d_{5/2}$  & 25.107 & 0.77(12)  & -0.79(13) & \cite{stone2011} \\
\hline                   
48 & 111 & 0.396   & 11/2- & $0h_{11/2}$ & 28.771 & -0.747(4) & 0.50(2)   & \cite{yordanov2013} \\
\hline                   
48 & 113 & 0.264   & 11/2- & $0h_{11/2}$ & 29.076 & -0.612(3) & 0.34(1)   & \cite{yordanov2013} \\
\hline                   
48 & 115 & 0.181   & 11/2- & $0h_{11/2}$ & 29.387 & -0.476(5) & 0.23(1)   & \cite{yordanov2013} \\
\hline                   
48 & 117 & 0.136   & 11/2- & $0h_{11/2}$ & 29.687 & -0.320(6) & 0.14(1)   & \cite{yordanov2013} \\
\hline                   
48 & 119 & 0.147   & 11/2- & $0h_{11/2}$ & 29.988 & -0.135(3) & 0.06(1)   & \cite{yordanov2013} \\
\hline                   
48 & 121 & 0       & 3/2+  & $1d_{3/2}$  & 27.671 & -0.274(7) & 0.33(3)   & \cite{yordanov2013} \\
\hline                   
48 & 121 & 0.215   & 11/2- & $0h_{11/2}$ & 30.285 & 0.009(6)  & -0.005(3) & \cite{yordanov2013} \\
\hline                   
48 & 123 & 0       & 3/2+  & $1d_{3/2}$  & 27.889 & 0.042(5)  & -0.06(1)  & \cite{yordanov2013} \\
\hline                    
48 & 123 & 0.317   & 11/2- & $0h_{11/2}$ & 30.587 & 0.135(4)  & -0.10(1)  & \cite{yordanov2013} \\
\hline                   
48 & 125 & 0       & 3/2+  & $1d_{3/2}$  & 28.092 & 0.209(4)  & -0.43(4)  & \cite{yordanov2013} \\
\hline                   
48 & 125 & 0+x     & 11/2- & $0h_{11/2}$ & 30.876 & 0.269(7)  & -0.26(3)  & \cite{yordanov2013} \\
\hline                   
48 & 127 & 0       & 3/2+  & $1d_{3/2}$  & 28.314 & 0.239(5)  & -0.72(6)  & \cite{yordanov2013} \\
\hline                   
48 & 127 & 0+x     & 11/2- & $0h_{11/2}$ & 31.170 & 0.342(10) & -0.49(4)  & \cite{yordanov2013} \\
\hline                   
48 & 129 & 0       & 3/2+  & $1d_{3/2}$  & 28.970 & 0.132(7)  & -0.56(6)  & \cite{yordanov2013} \\
\hline                   
48 & 129 & 0+x     & 11/2- & $0h_{11/2}$ & 31.455 & 0.570(13) & -1.16(11) & \cite{yordanov2013} \\
\hline                   
50 & 109 & 0       & 5/2+  & $1d_{5/2}$  & 24.358 & 0.31(10)  & -1.09(36) & \cite{stone2011} \\
\hline                   
50 & 111 & 0       & 7/2+  & $0g_{7/2}$  & 24.826 & 0.18(9)   & -0.52(26) & \cite{stone2011} \\
\hline                   
50 & 113 & 0.738   & 11/2- & $0h_{11/2}$ & 28.754 & -0.41(4)  & 0.91(9)   & \cite{stone2011} \\
\hline                   
50 & 115 & 0.613   & 7/2+  & $0g_{7/2}$  & 25.357 & 0.26(3)   & -0.80(9)  & \cite{stone2011} \\
\hline                   
50 & 115 & 0.714   & 11/2- & $0h_{11/2}$ & 29.061 & -0.38(6)  & 0.89(14)  & \cite{stone2011} \\
\hline                   
50 & 117 & 0.315   & 11/2- & $0h_{11/2}$ & 29.365 & -0.42(5)  & 0.99(12)  & \cite{stone2011} \\
\hline                   
50 & 119 & 0.024   & 3/2+  & $1d_{3/2}$  & 26.833 & -0.112(7) & 0.54(4)   & \cite{stone2011} \\
\hline                   
50 & 119 & 0.09    & 11/2- & $0h_{11/2}$ & 29.662 & -0.21(2)  & 0.48(5)   & \cite{stone2011} \\
\hline                   
50 & 121 & 0       & 3/2+  & $1d_{3/2}$  & 27.058 & -0.02(2)  & 0.10(10)  & \cite{stone2011} \\
\hline                   
50 & 121 & 0.006   & 11/2- & $0h_{11/2}$ & 29.967 & -0.14(3)  & 0.32(7)   & \cite{stone2011} \\
\hline                   
50 & 123 & 0       & 11/2- & $0h_{11/2}$ & 30.257 & 0.03(4)   & -0.07(10) & \cite{stone2011} \\
\hline                   
50 & 125 & 0       & 11/2- & $0h_{11/2}$ & 30.555 & 0.14(21)  & -0.38(57) & \cite{leblanc2005} \\
\hline                   
50 & 127 & 0       & 11/2- & $0h_{11/2}$ & 30.848 & 0.30(13)  & -1.01(44) & \cite{leblanc2005} \\
\hline                   
50 & 129 & 0.035   & 11/2- & $0h_{11/2}$ & 31.136 & -0.18(17) & 0.87(82)  & \cite{leblanc2005} \\
\hline                   
50 & 131 & 0+x     & 11/2- & $0h_{11/2}$ & 31.423 & 0.02(20)  & -0.12(120) & \cite{leblanc2005} \\
\hline                   
\end{tabular}
\caption{\label{tab:2/tc} Experimental quadrupole moments, and deduced occupation functions $(u^2-v^2)$ of a sequence of Cd and Sn isotopes for ${\bar \chi} = 0.3\times 10^{-3} MeV-fm^{-4}$.
For $^{113}\text{Sn}(11/-)$, $^{115}\text{Sn}(7/2+)$, $^{115}\text{Sn}(11/2-)$, $^{119}\text{Sn}(11/2-)$ the signs shown in the table are based on systematics since only $ |Q_s| $ is known.}

\end{table}

\begin{figure}[h]
\centering
\includegraphics[scale=0.5 ]{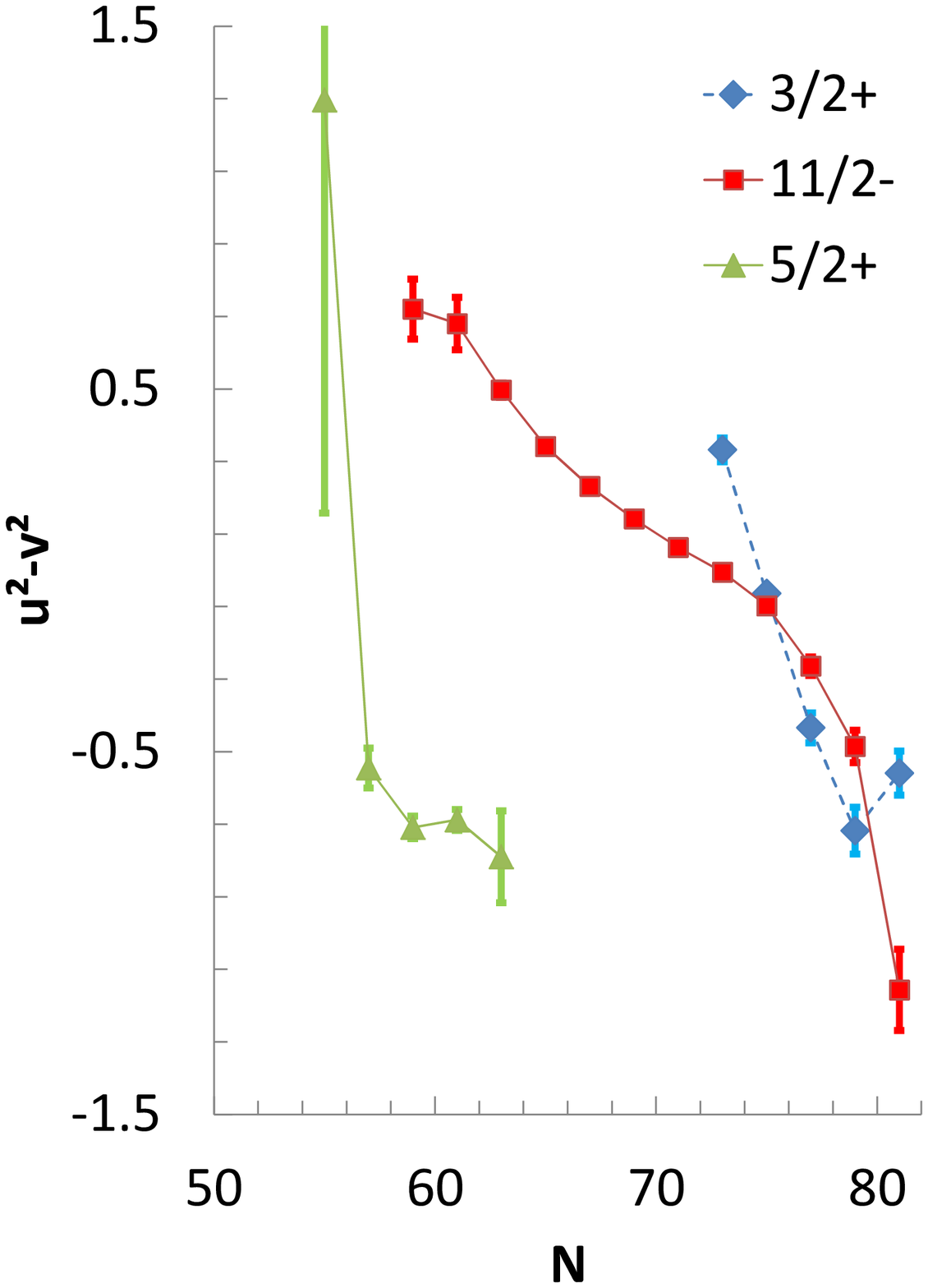}
\caption{\label{fig:cd-u2-v2}(Color online) Occupation function $(u^2-v^2)$ for the Cd isotopes for ${ \bar \chi =0.3 \times 10^{-3} }$.}
\end{figure}

\begin{figure}[h]
\centering
\includegraphics[scale=0.5 ]{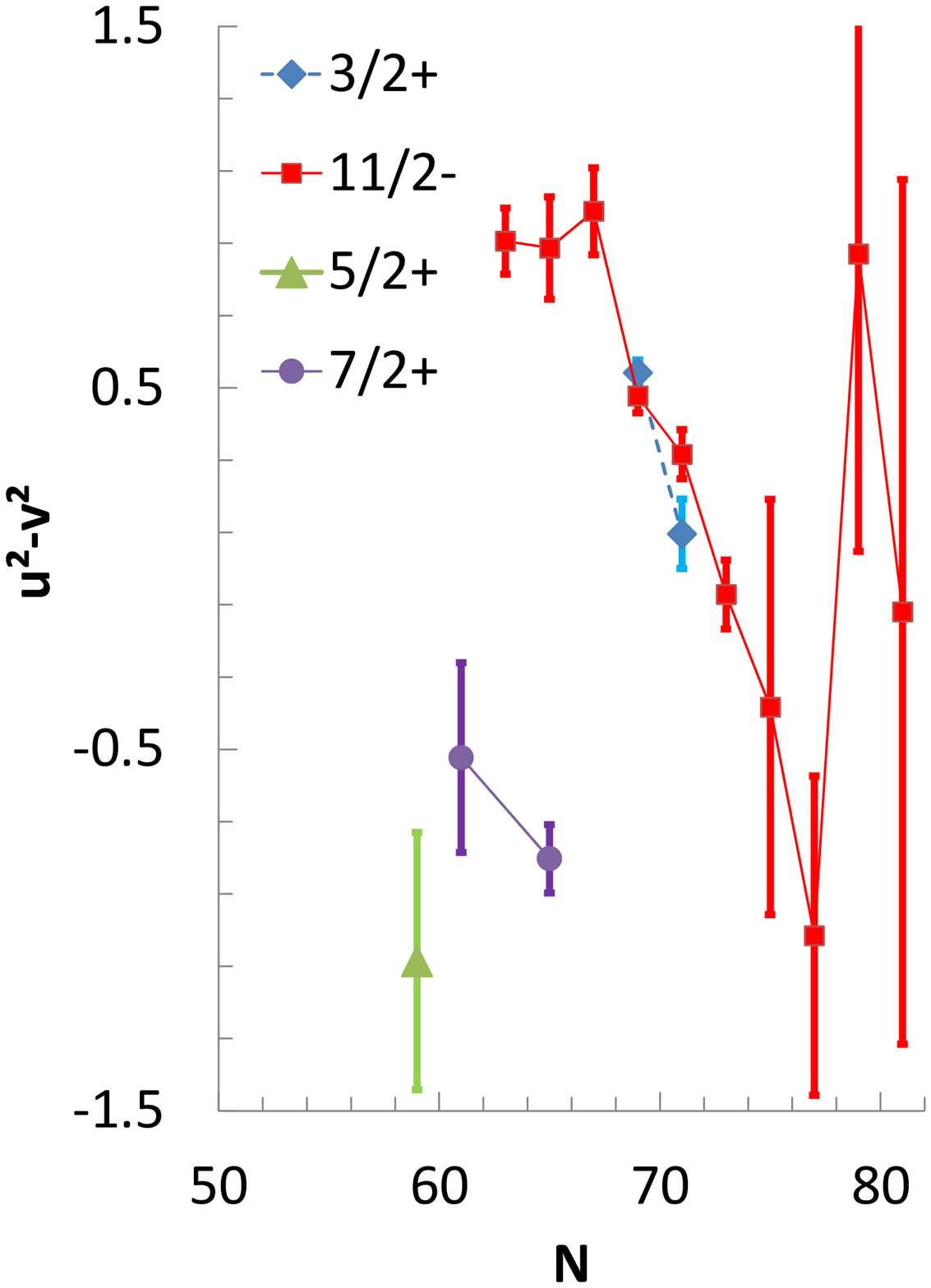}
\caption{\label{fig:sn-u2-v2}(Color online) Occupation function $(u^2-v^2)$ for the Sn isotopes for ${\bar \chi =0.3 \times 10^{-3}}$.}
\end{figure}

The extracted occupation probabilities are consistent with expectations with the exception of $^{129}\text{Sn}$. Since the $h_{11/2}$ orbital cannot be less than half filled in $^{129}\text{Sn}$, the measured negative quadrupole moment of the $11/2-$  level, even with a very large error bar, would imply a major change in structure compared to the other Sn isotopes
 but there is no other evidence for this happening.Leaving out $^{129}\text{Sn}$, the total effective charge $e_{n,tot}/e$ of the $h_{11/2}$ quasiparticle states in the Sn isotopes, 
varies relatively weakly, ranging from $1.2$ to $2.0$, and the occupation probabilities are therefore a roughly linear function of the neutron number N. This is very similar to the model
advocated by Yordanov et al. \cite{yordanov2013}. However, the Cd isotopes are more collective as evidenced by the lower energy and larger $B(E2)$ of the $2+$ states and 
consequently the $h_{11/2}$ total effective charge $e_{n,tot}/e$ is much larger and more variable, ranging from $2.5$ to $9.5$ near mid-shell. The occupation probabilities have a more
 complicated N dependence than assumed by Yordanov et al. \cite{yordanov2013}  despite the remarkably linear N dependence of the quadrupole moments. 

It is interesting that the data for the $d_{5/2}$ quasi-particle state in the Cd isotopes clearly shows that this orbital fills more rapidly that the $g_{7/2}$ orbital.
The Sn data are consistent with this.

It is a limitation of laser spectroscopy that it can provide information only on the properties of the ground state and isomeric states. But a great strength is that it can do so
 with a consistent methodology for long chains of isotopes. Since occupation probablities are monotonic though non-linear functions of N in pretty much all models, 
the data on each isotope propagates information along the whole chain.For example, in the Cd isotopes the occupation probability of the $d_{5/2}$ orbital is close to $0.9$
by N=63 and this value will only increase with increasing N; similarly, the occupation probability of the $d_{3/2}$ orbital is down to $0.3$ by $N=73$ and will only decrease with 
decreasing N.

Since the coefficients $|C_2|^2$ turn out to be small (always less than 0.03 for the Sn isotopes and less than 0.15 for the Cd isotopes and much less when $|u^2-v^2|$ is small ), 
the weak coupling approximation is reasonable in the present case. However, calculating the quadrupole moments to first order in $C_2$ also assumes that the quadrupole moment
of the collective $2+$ state in the even isotopes is not too large. Fortunately, this is so. Where experimental data are available, $Q(2+)$ ranges from -0.28 b to -0.45 b in Cd and
 is substantially smaller in Sn. The resulting correction to the first order calculation would be less than $6\%$ in the Cd isotopes
 and less than $2\%$ in the Sn isotopes .

The value of ${\bar \chi}$ has a significant uncertainty but it doesn't affect the shape of the curves in Figs.1 and 2 though it controls the scale of the vertical axis. 
The fact that a constant value works well for both isotope chains provides evidence for the validity of the approximation that $\chi_{nn} / \chi_{pn} \approx 1/3$ since the 
$2+$ state is expected to be mainly a two neutron quasi-particle state in Sn but a mixed proton and neutron two quasi-particle state in Cd. However, the noticeable discontinuity 
at $^{129}\text{Cd}$ just below the $N=82$ neutron shell closure may indicate the limitation of the approximation. It may also be noted that the self-consistent estimate  \cite{bm-vol2-1975,dobaczewski1988, bes-sorensen-1995}  of the quadrupole interaction strengths gives $\chi_{nn}=79 A^{-7/3}$ and $\chi_{pn}=387 A^{-7/3}$. 
Our value is smaller than the self consistent $\chi_{nn}$ by a factor of roughly four, but this is perhaps not surprising since it multiplies an experimental $B(E2)$ in equ.~\ref{q_moment} which can include various renormalization effects.

\section{conclusions}

I have shown that the measured electic quadrupole moments of a long chain of odd Cd and Sn isotopes can be simply understood in terms of a weak coupling model. 
In the model, the quadrupole moments are proportional to the occupation probability of the relevant orbitals which can therefore be extracted from the data. 
The formula for the quadrupole moment requires the empirical determination one strength parameter ( which is held constant for the two isotope chains), 
but otherwise it involves experimental data and standard values for parameters such as effective charges. 

The deduced occupation probabilities agree with expectations except for $^{129}\text{Sn}$. It would be interesting to repeat the experiments on the heavy Sn isotopes near the 
doubly closed shell $^{132}\text{Sn}$ in order to decrease the large error bars for $Q_s(h_{11/2})$ and to measure $Q_s(d_{3/2})$.

\section{acknowledgments.}
Discussions with J. Crawford and F. Buchinger are gratefully acknowledged. 
This work was supported by an internal research grant from McGill University. 

\clearpage 


\begin{thebibliography}{20}%
\makeatletter
\providecommand \@ifxundefined [1]{%
 \@ifx{#1\undefined}
}%
\providecommand \@ifnum [1]{%
 \ifnum #1\expandafter \@firstoftwo
 \else \expandafter \@secondoftwo
 \fi
}%
\providecommand \@ifx [1]{%
 \ifx #1\expandafter \@firstoftwo
 \else \expandafter \@secondoftwo
 \fi
}%
\providecommand \natexlab [1]{#1}%
\providecommand \enquote  [1]{``#1''}%
\providecommand \bibnamefont  [1]{#1}%
\providecommand \bibfnamefont [1]{#1}%
\providecommand \citenamefont [1]{#1}%
\providecommand \href@noop [0]{\@secondoftwo}%
\providecommand \href [0]{\begingroup \@sanitize@url \@href}%
\providecommand \@href[1]{\@@startlink{#1}\@@href}%
\providecommand \@@href[1]{\endgroup#1\@@endlink}%
\providecommand \@sanitize@url [0]{\catcode `\\12\catcode `\$12\catcode
  `\&12\catcode `\#12\catcode `\^12\catcode `\_12\catcode `\%12\relax}%
\providecommand \@@startlink[1]{}%
\providecommand \@@endlink[0]{}%
\providecommand \url  [0]{\begingroup\@sanitize@url \@url }%
\providecommand \@url [1]{\endgroup\@href {#1}{\urlprefix }}%
\providecommand \urlprefix  [0]{URL }%
\providecommand \Eprint [0]{\href }%
\providecommand \doibase [0]{http://dx.doi.org/}%
\providecommand \selectlanguage [0]{\@gobble}%
\providecommand \bibinfo  [0]{\@secondoftwo}%
\providecommand \bibfield  [0]{\@secondoftwo}%
\providecommand \translation [1]{[#1]}%
\providecommand \BibitemOpen [0]{}%
\providecommand \bibitemStop [0]{}%
\providecommand \bibitemNoStop [0]{.\EOS\space}%
\providecommand \EOS [0]{\spacefactor3000\relax}%
\providecommand \BibitemShut  [1]{\csname bibitem#1\endcsname}%
\let\auto@bib@innerbib\@empty
\bibitem [{\citenamefont {Yordanov}\ \emph {et~al.}(2013)\citenamefont
  {Yordanov}, \citenamefont {Balabanski}, \citenamefont
  {Biero\ifmmode~\acute{n}\else \'{n}\fi{}}, \citenamefont {Bissell},
  \citenamefont {Blaum}, \citenamefont {Budincevic}, \citenamefont {Fritzsche},
  \citenamefont {Fr\"ommgen}, \citenamefont {Georgiev}, \citenamefont
  {Geppert}, \citenamefont {Hammen}, \citenamefont {Kowalska}, \citenamefont
  {Kreim}, \citenamefont {Krieger}, \citenamefont {Neugart}, \citenamefont
  {N\"ortersh\"auser}, \citenamefont {Papuga},\ and\ \citenamefont
  {Schmidt}}]{yordanov2013}%
  \BibitemOpen
  \bibfield  {author} {\bibinfo {author} {\bibfnamefont {D.~T.}\ \bibnamefont
  {Yordanov}}, \bibinfo {author} {\bibfnamefont {D.~L.}\ \bibnamefont
  {Balabanski}}, \bibinfo {author} {\bibfnamefont {J.}~\bibnamefont
  {Biero\ifmmode~\acute{n}\else \'{n}\fi{}}}, \bibinfo {author} {\bibfnamefont
  {M.~L.}\ \bibnamefont {Bissell}}, \bibinfo {author} {\bibfnamefont
  {K.}~\bibnamefont {Blaum}}, \bibinfo {author} {\bibfnamefont
  {I.}~\bibnamefont {Budincevic}}, \bibinfo {author} {\bibfnamefont
  {S.}~\bibnamefont {Fritzsche}}, \bibinfo {author} {\bibfnamefont
  {N.}~\bibnamefont {Fr\"ommgen}}, \bibinfo {author} {\bibfnamefont
  {G.}~\bibnamefont {Georgiev}}, \bibinfo {author} {\bibfnamefont
  {C.}~\bibnamefont {Geppert}}, \bibinfo {author} {\bibfnamefont
  {M.}~\bibnamefont {Hammen}}, \bibinfo {author} {\bibfnamefont
  {M.}~\bibnamefont {Kowalska}}, \bibinfo {author} {\bibfnamefont
  {K.}~\bibnamefont {Kreim}}, \bibinfo {author} {\bibfnamefont
  {A.}~\bibnamefont {Krieger}}, \bibinfo {author} {\bibfnamefont
  {R.}~\bibnamefont {Neugart}}, \bibinfo {author} {\bibfnamefont
  {W.}~\bibnamefont {N\"ortersh\"auser}}, \bibinfo {author} {\bibfnamefont
  {J.}~\bibnamefont {Papuga}}, \ and\ \bibinfo {author} {\bibfnamefont
  {S.}~\bibnamefont {Schmidt}},\ }\href@noop {} {\bibfield  {journal} {\bibinfo
   {journal} {Phys. Rev. Lett.}\ }\textbf {\bibinfo {volume} {110}},\ \bibinfo
  {pages} {192501} (\bibinfo {year} {2013})}\BibitemShut {NoStop}%
\bibitem [{\citenamefont {Ring}\ and\ \citenamefont
  {Schuck}(1980)}]{ring-schuck-1980}%
  \BibitemOpen
  \bibfield  {author} {\bibinfo {author} {\bibfnamefont {P.}~\bibnamefont
  {Ring}}\ and\ \bibinfo {author} {\bibfnamefont {P.}~\bibnamefont {Schuck}},\
  }\href@noop {} {\emph {\bibinfo {title} {The Nuclear Many-Body Problem}}}\
  (\bibinfo  {publisher} {Springer},\ \bibinfo {year} {1980})\BibitemShut
  {NoStop}%
\bibitem [{\citenamefont {Bes}\ and\ \citenamefont
  {Sorensen}(1995)}]{bes-sorensen-1995}%
  \BibitemOpen
  \bibfield  {author} {\bibinfo {author} {\bibfnamefont {D.~R.}\ \bibnamefont
  {Bes}}\ and\ \bibinfo {author} {\bibfnamefont {R.~A.}\ \bibnamefont
  {Sorensen}},\ }\href@noop {} {\bibfield  {journal} {\bibinfo  {journal}
  {Adv.Nucl.Phys.}\ }\textbf {\bibinfo {volume} {2}},\ \bibinfo {pages} {129 }
  (\bibinfo {year} {1995})}\BibitemShut {NoStop}%
\bibitem [{bro(2001)}]{brown2001}%
  \BibitemOpen
  \bibfield  {author} {\bibinfo {author} {\bibfnamefont {B.~A.}\ \bibnamefont {Brown}},\ }\href@noop{}
  {\bibfield  {journal} {\bibinfo  {journal} {Progress in
  Particle and Nuclear Physics}\ }\textbf {\bibinfo {volume} {47}},\ \bibinfo
  {pages} {517 } (\bibinfo {year} {2001})}\BibitemShut {NoStop}%
\bibitem [{\citenamefont {Honma}\ \emph {et~al.}(2004)\citenamefont {Honma},
  \citenamefont {Otsuka}, \citenamefont {Brown},\ and\ \citenamefont
  {Mizusaki}}]{honma2004}%
  \BibitemOpen
  \bibfield  {author} {\bibinfo {author} {\bibfnamefont {M.}~\bibnamefont
  {Honma}}, \bibinfo {author} {\bibfnamefont {T.}~\bibnamefont {Otsuka}},
  \bibinfo {author} {\bibfnamefont {B.~A.}\ \bibnamefont {Brown}}, \ and\
  \bibinfo {author} {\bibfnamefont {T.}~\bibnamefont {Mizusaki}},\ }\href@noop
  {} 
  {\bibfield  {journal} {\bibinfo  {journal} {Phys. Rev. C}\ }\textbf
  {\bibinfo {volume} {69}},\ \bibinfo {pages} {034335} (\bibinfo {year}
  {2004})}\BibitemShut {NoStop}%
\bibitem [{\citenamefont {Terasaki}\ \emph {et~al.}(2002)\citenamefont
  {Terasaki}, \citenamefont {Engel}, \citenamefont {Nazarewicz},\ and\
  \citenamefont {Stoitsov}}]{terasaki2002}%
  \BibitemOpen
  \bibfield  {author} {\bibinfo {author} {\bibfnamefont {J.}~\bibnamefont
  {Terasaki}}, \bibinfo {author} {\bibfnamefont {J.}~\bibnamefont {Engel}},
  \bibinfo {author} {\bibfnamefont {W.}~\bibnamefont {Nazarewicz}}, \ and\
  \bibinfo {author} {\bibfnamefont {M.}~\bibnamefont {Stoitsov}},\ }\href@noop
  {} {\bibfield  {journal} {\bibinfo  {journal} {Phys. Rev. C}\ }\textbf
  {\bibinfo {volume} {66}},\ \bibinfo {pages} {054313} (\bibinfo {year}
  {2002})}\BibitemShut {NoStop}%
\bibitem [{\citenamefont {Dobaczewski}\ \emph {et~al.}(1988)\citenamefont
  {Dobaczewski}, \citenamefont {Nazarewicz}, \citenamefont {Skalski},\ and\
  \citenamefont {Werner}}]{dobaczewski1988}%
  \BibitemOpen
  \bibfield  {author} {\bibinfo {author} {\bibfnamefont {J.}~\bibnamefont
  {Dobaczewski}}, \bibinfo {author} {\bibfnamefont {W.}~\bibnamefont
  {Nazarewicz}}, \bibinfo {author} {\bibfnamefont {J.}~\bibnamefont {Skalski}},
  \ and\ \bibinfo {author} {\bibfnamefont {T.}~\bibnamefont {Werner}},\
  }\href@noop {} {\bibfield  {journal} {\bibinfo  {journal} {Phys. Rev. Lett.}\
  }\textbf {\bibinfo {volume} {60}},\ \bibinfo {pages} {2254} (\bibinfo {year}
  {1988})}\BibitemShut {NoStop}%
\bibitem [{\citenamefont {Lojewski}\ \emph {et~al.}(2001)\citenamefont
  {Lojewski}, \citenamefont {Nerlo-Pomorska},\ and\ \citenamefont
  {Dudek}}]{lojewski2001}%
  \BibitemOpen
  \bibfield  {author} {\bibinfo {author} {\bibfnamefont {Z.}~\bibnamefont
  {Lojewski}}, \bibinfo {author} {\bibfnamefont {B.}~\bibnamefont
  {Nerlo-Pomorska}}, \ and\ \bibinfo {author} {\bibfnamefont {J.}~\bibnamefont
  {Dudek}},\ }\href@noop {} {\bibfield  {journal} {\bibinfo  {journal} {Acta
  Phys. Pol. B}\ }\textbf {\bibinfo {volume} {32}},\ \bibinfo {pages} {2981}
  (\bibinfo {year} {2001})}\BibitemShut {NoStop}%
\bibitem [{\citenamefont {Boelaert}\ \emph {et~al.}(2007)\citenamefont
  {Boelaert}, \citenamefont {Dewald}, \citenamefont {Fransen}, \citenamefont
  {Jolie}, \citenamefont {Linnemann}, \citenamefont {Melon}, \citenamefont
  {M\"oller}, \citenamefont {Smirnova},\ and\ \citenamefont
  {Heyde}}]{boelaert2007}%
  \BibitemOpen
  \bibfield  {author} {\bibinfo {author} {\bibfnamefont {N.}~\bibnamefont
  {Boelaert}}, \bibinfo {author} {\bibfnamefont {A.}~\bibnamefont {Dewald}},
  \bibinfo {author} {\bibfnamefont {C.}~\bibnamefont {Fransen}}, \bibinfo
  {author} {\bibfnamefont {J.}~\bibnamefont {Jolie}}, \bibinfo {author}
  {\bibfnamefont {A.}~\bibnamefont {Linnemann}}, \bibinfo {author}
  {\bibfnamefont {B.}~\bibnamefont {Melon}}, \bibinfo {author} {\bibfnamefont
  {O.}~\bibnamefont {M\"oller}}, \bibinfo {author} {\bibfnamefont
  {N.}~\bibnamefont {Smirnova}}, \ and\ \bibinfo {author} {\bibfnamefont
  {K.}~\bibnamefont {Heyde}},\ }\href@noop {} {\bibfield  {journal} {\bibinfo
  {journal} {Phys. Rev. C}\ }\textbf {\bibinfo {volume} {75}},\ \bibinfo
  {pages} {054311} (\bibinfo {year} {2007})}\BibitemShut {NoStop}%
\bibitem [{\citenamefont {Doornenbal}\ \emph {et~al.}()\citenamefont
  {Doornenbal}, \citenamefont {Takeuchi}, \citenamefont {Aoi}, \citenamefont
  {Matsushita}, \citenamefont {Obertelli}, \citenamefont {Steppenbeck},
  \citenamefont {Wang}, \citenamefont {Audirac}, \citenamefont {Baba},
  \citenamefont {P.Bednarczyk}, \citenamefont {Boissinot}, \citenamefont
  {Ciemala}, \citenamefont {Corsi}, \citenamefont {Furumoto}, \citenamefont
  {Isobe}, \citenamefont {Jungclaus}, \citenamefont {Lapoux}, \citenamefont
  {Lee}, \citenamefont {Matsui}, \citenamefont {Motobayashi}, \citenamefont
  {Nishimura}, \citenamefont {Ota}, \citenamefont {Pollacco}, \citenamefont
  {Sakurai}, \citenamefont {Santamaria}, \citenamefont {Shiga}, \citenamefont
  {Sohler},\ and\ \citenamefont {Taniuchi}}]{doornenbal2013}%
  \BibitemOpen
  \bibfield  {author} {\bibinfo {author} {\bibfnamefont {P.}~\bibnamefont
  {Doornenbal}}, \bibinfo {author} {\bibfnamefont {S.}~\bibnamefont
  {Takeuchi}}, \bibinfo {author} {\bibfnamefont {N.}~\bibnamefont {Aoi}},
  \bibinfo {author} {\bibfnamefont {M.}~\bibnamefont {Matsushita}}, \bibinfo
  {author} {\bibfnamefont {A.}~\bibnamefont {Obertelli}}, \bibinfo {author}
  {\bibfnamefont {D.}~\bibnamefont {Steppenbeck}}, \bibinfo {author}
  {\bibfnamefont {H.}~\bibnamefont {Wang}}, \bibinfo {author} {\bibfnamefont
  {L.}~\bibnamefont {Audirac}}, \bibinfo {author} {\bibfnamefont
  {H.}~\bibnamefont {Baba}}, \bibinfo {author} {\bibnamefont {P.Bednarczyk}},
  \bibinfo {author} {\bibfnamefont {S.}~\bibnamefont {Boissinot}}, \bibinfo
  {author} {\bibfnamefont {M.}~\bibnamefont {Ciemala}}, \bibinfo {author}
  {\bibfnamefont {A.}~\bibnamefont {Corsi}}, \bibinfo {author} {\bibfnamefont
  {T.}~\bibnamefont {Furumoto}}, \bibinfo {author} {\bibfnamefont
  {T.}~\bibnamefont {Isobe}}, \bibinfo {author} {\bibfnamefont
  {A.}~\bibnamefont {Jungclaus}}, \bibinfo {author} {\bibfnamefont
  {V.}~\bibnamefont {Lapoux}}, \bibinfo {author} {\bibfnamefont
  {J.}~\bibnamefont {Lee}}, \bibinfo {author} {\bibfnamefont {K.}~\bibnamefont
  {Matsui}}, \bibinfo {author} {\bibfnamefont {T.}~\bibnamefont {Motobayashi}},
  \bibinfo {author} {\bibfnamefont {D.}~\bibnamefont {Nishimura}}, \bibinfo
  {author} {\bibfnamefont {S.}~\bibnamefont {Ota}}, \bibinfo {author}
  {\bibfnamefont {E.}~\bibnamefont {Pollacco}}, \bibinfo {author}
  {\bibfnamefont {H.}~\bibnamefont {Sakurai}}, \bibinfo {author} {\bibfnamefont
  {C.}~\bibnamefont {Santamaria}}, \bibinfo {author} {\bibfnamefont
  {Y.}~\bibnamefont {Shiga}}, \bibinfo {author} {\bibfnamefont
  {D.}~\bibnamefont {Sohler}}, \ and\ \bibinfo {author} {\bibfnamefont
  {R.}~\bibnamefont {Taniuchi}},\ }\href@noop {} {\bibinfo  {journal}
  {arXiv:1305.2877v1 [nucl-ex]}\ }\BibitemShut {NoStop}%
\bibitem [{\citenamefont {Kumar}\ \emph {et~al.}(2010)\citenamefont {Kumar},
  \citenamefont {Doornenbal}, \citenamefont {Jhingan}, \citenamefont {Bhowmik},
  \citenamefont {Muralithar}, \citenamefont {Appannababu}, \citenamefont
  {Garg}, \citenamefont {Gerl}, \citenamefont {Gorska}, \citenamefont {Kaur},
  \citenamefont {Kojouharov}, \citenamefont {Mandal}, \citenamefont
  {Mukherjee}, \citenamefont {Siwal}, \citenamefont {Sharma}, \citenamefont
  {Singh}, \citenamefont {Singh},\ and\ \citenamefont
  {Wollersheim}}]{kumar2010}%
  \BibitemOpen
\bibfield  {journal} {  }\bibfield  {author} {\bibinfo {author} {\bibfnamefont
  {R.}~\bibnamefont {Kumar}}, \bibinfo {author} {\bibfnamefont
  {P.}~\bibnamefont {Doornenbal}}, \bibinfo {author} {\bibfnamefont
  {A.}~\bibnamefont {Jhingan}}, \bibinfo {author} {\bibfnamefont {R.~K.}\
  \bibnamefont {Bhowmik}}, \bibinfo {author} {\bibfnamefont {S.}~\bibnamefont
  {Muralithar}}, \bibinfo {author} {\bibfnamefont {S.}~\bibnamefont
  {Appannababu}}, \bibinfo {author} {\bibfnamefont {R.}~\bibnamefont {Garg}},
  \bibinfo {author} {\bibfnamefont {J.}~\bibnamefont {Gerl}}, \bibinfo {author}
  {\bibfnamefont {M.}~\bibnamefont {Gorska}}, \bibinfo {author} {\bibfnamefont
  {J.}~\bibnamefont {Kaur}}, \bibinfo {author} {\bibfnamefont {I.}~\bibnamefont
  {Kojouharov}}, \bibinfo {author} {\bibfnamefont {S.}~\bibnamefont {Mandal}},
  \bibinfo {author} {\bibfnamefont {S.}~\bibnamefont {Mukherjee}}, \bibinfo
  {author} {\bibfnamefont {D.}~\bibnamefont {Siwal}}, \bibinfo {author}
  {\bibfnamefont {A.}~\bibnamefont {Sharma}}, \bibinfo {author} {\bibfnamefont
  {P.~P.}\ \bibnamefont {Singh}}, \bibinfo {author} {\bibfnamefont {R.~P.}\
  \bibnamefont {Singh}}, \ and\ \bibinfo {author} {\bibfnamefont {H.~J.}\
  \bibnamefont {Wollersheim}},\ }\href@noop {} {\bibfield  {journal} {\bibinfo
  {journal} {Phys. Rev. C}\ }\textbf {\bibinfo {volume} {81}},\ \bibinfo
  {pages} {024306} (\bibinfo {year} {2010})}\BibitemShut {NoStop}%
\bibitem [{\citenamefont {Raman}\ \emph {et~al.}(2001)\citenamefont {Raman},
  \citenamefont {JR.},\ and\ \citenamefont {Tikkanen}}]{raman2001}%
  \BibitemOpen
  \bibfield  {author} 
  {\bibinfo {author} {\bibfnamefont {S.}~\bibnamefont {Raman}}, 
   \bibinfo {author} {\bibfnamefont {C.~N.}~\bibnamefont {Nestor}\ \bibnamefont {JR.}}, \
  and\ \bibinfo {author} {\bibfnamefont {P.}~\bibnamefont {Tikkanen}},\
  }\href@noop {} {\bibfield  {journal} {\bibinfo  {journal} {Atomic Data and
  Nuclear Data Tables}\ }\textbf {\bibinfo {volume} {78}},\ \bibinfo {pages} {1
  } (\bibinfo {year} {2001})}\BibitemShut {NoStop}%
\bibitem [{\citenamefont {Ilieva}\ \emph {et~al.}(2013)\citenamefont {Ilieva},
  \citenamefont {Bonig}, \citenamefont {Hartig}, \citenamefont {Henrich},
  \citenamefont {Ignatov}, \citenamefont {Kroll}, \citenamefont {Seiert},
  \citenamefont {Thurauf}, \citenamefont {Duppen}, \citenamefont {Huyse},
  \citenamefont {Raabe}, \citenamefont {Sotty}, \citenamefont {Witte},
  \citenamefont {Gottberg}, \citenamefont {Mendonca}, \citenamefont
  {Rapisarda}, \citenamefont {Blazhev}, \citenamefont {Jolie}, \citenamefont
  {Regis}, \citenamefont {Saed-Samii}, \citenamefont {Warr}, \citenamefont
  {Filipescu}, \citenamefont {Gheorghe}, \citenamefont {Ghita}, \citenamefont
  {Lica}, \citenamefont {Marginean}, \citenamefont {Koster}, \citenamefont
  {Thirolf}, \citenamefont {Fraile}, \citenamefont {Paziy}, \citenamefont
  {Vedia}, \citenamefont {Andreyev}, \citenamefont {Algora}, \citenamefont
  {Simpson}, \citenamefont {Podolyak}, \citenamefont {Mach},\ and\
  \citenamefont {Cheal}}]{ilieva2013}%
  \BibitemOpen
  \bibfield  {author} {\bibinfo {author} {\bibfnamefont {S.}~\bibnamefont
  {Ilieva}}, \bibinfo {author} {\bibfnamefont {S.}~\bibnamefont {Bonig}},
  \bibinfo {author} {\bibfnamefont {A.-L.}\ \bibnamefont {Hartig}}, \bibinfo
  {author} {\bibfnamefont {C.}~\bibnamefont {Henrich}}, \bibinfo {author}
  {\bibfnamefont {A.}~\bibnamefont {Ignatov}}, \bibinfo {author} {\bibfnamefont
  {T.}~\bibnamefont {Kroll}}, \bibinfo {author} {\bibfnamefont
  {C.}~\bibnamefont {Seiert}}, \bibinfo {author} {\bibfnamefont
  {M.}~\bibnamefont {Thurauf}}, \bibinfo {author} {\bibfnamefont {P.~V.}\
  \bibnamefont {Duppen}}, \bibinfo {author} {\bibfnamefont {M.}~\bibnamefont
  {Huyse}}, \bibinfo {author} {\bibfnamefont {R.}~\bibnamefont {Raabe}},
  \bibinfo {author} {\bibfnamefont {C.}~\bibnamefont {Sotty}}, \bibinfo
  {author} {\bibfnamefont {H.~D.}\ \bibnamefont {Witte}}, \bibinfo {author}
  {\bibfnamefont {A.}~\bibnamefont {Gottberg}}, \bibinfo {author}
  {\bibfnamefont {T.}~\bibnamefont {Mendonca}}, \bibinfo {author}
  {\bibfnamefont {E.}~\bibnamefont {Rapisarda}}, \bibinfo {author}
  {\bibfnamefont {A.}~\bibnamefont {Blazhev}}, \bibinfo {author} {\bibfnamefont
  {J.}~\bibnamefont {Jolie}}, \bibinfo {author} {\bibfnamefont {J.-M.}\
  \bibnamefont {Regis}}, \bibinfo {author} {\bibfnamefont {N.}~\bibnamefont
  {Saed-Samii}}, \bibinfo {author} {\bibfnamefont {N.}~\bibnamefont {Warr}},
  \bibinfo {author} {\bibfnamefont {D.}~\bibnamefont {Filipescu}}, \bibinfo
  {author} {\bibfnamefont {I.}~\bibnamefont {Gheorghe}}, \bibinfo {author}
  {\bibfnamefont {D.}~\bibnamefont {Ghita}}, \bibinfo {author} {\bibfnamefont
  {R.}~\bibnamefont {Lica}}, \bibinfo {author} {\bibfnamefont {N.}~\bibnamefont
  {Marginean}}, \bibinfo {author} {\bibfnamefont {U.}~\bibnamefont {Koster}},
  \bibinfo {author} {\bibfnamefont {P.}~\bibnamefont {Thirolf}}, \bibinfo
  {author} {\bibfnamefont {L.}~\bibnamefont {Fraile}}, \bibinfo {author}
  {\bibfnamefont {V.}~\bibnamefont {Paziy}}, \bibinfo {author} {\bibfnamefont
  {V.}~\bibnamefont {Vedia}}, \bibinfo {author} {\bibfnamefont
  {A.}~\bibnamefont {Andreyev}}, \bibinfo {author} {\bibfnamefont
  {A.}~\bibnamefont {Algora}}, \bibinfo {author} {\bibfnamefont
  {G.}~\bibnamefont {Simpson}}, \bibinfo {author} {\bibfnamefont
  {Z.}~\bibnamefont {Podolyak}}, \bibinfo {author} {\bibfnamefont
  {H.}~\bibnamefont {Mach}}, \ and\ \bibinfo {author} {\bibfnamefont
  {B.}~\bibnamefont {Cheal}},\ }\href
  {http://cds.cern.ch/record/1602593/files/INTC-P-383.pdf} {} (\bibinfo {year}
  {2013})\BibitemShut {NoStop}%
\bibitem [{\citenamefont {Allmond}\ \emph {et~al.}(2011)\citenamefont
  {Allmond}, \citenamefont {Radford}, \citenamefont {Baktash}, \citenamefont
  {Batchelder}, \citenamefont {Galindo-Uribarri}, \citenamefont {Gross},
  \citenamefont {Hausladen}, \citenamefont {Lagergren}, \citenamefont
  {Larochelle}, \citenamefont {Padilla-Rodal},\ and\ \citenamefont
  {Yu}}]{allmond2011}%
  \BibitemOpen
  \bibfield  {author} {\bibinfo {author} {\bibfnamefont {J.~M.}\ \bibnamefont
  {Allmond}}, \bibinfo {author} {\bibfnamefont {D.~C.}\ \bibnamefont
  {Radford}}, \bibinfo {author} {\bibfnamefont {C.}~\bibnamefont {Baktash}},
  \bibinfo {author} {\bibfnamefont {J.~C.}\ \bibnamefont {Batchelder}},
  \bibinfo {author} {\bibfnamefont {A.}~\bibnamefont {Galindo-Uribarri}},
  \bibinfo {author} {\bibfnamefont {C.~J.}\ \bibnamefont {Gross}}, \bibinfo
  {author} {\bibfnamefont {P.~A.}\ \bibnamefont {Hausladen}}, \bibinfo {author}
  {\bibfnamefont {K.}~\bibnamefont {Lagergren}}, \bibinfo {author}
  {\bibfnamefont {Y.}~\bibnamefont {Larochelle}}, \bibinfo {author}
  {\bibfnamefont {E.}~\bibnamefont {Padilla-Rodal}}, \ and\ \bibinfo {author}
  {\bibfnamefont {C.-H.}\ \bibnamefont {Yu}},\ }\href@noop {} {\bibfield
  {journal} {\bibinfo  {journal} {Phys. Rev. C}\ }\textbf {\bibinfo {volume}
  {84}},\ \bibinfo {pages} {061303} (\bibinfo {year} {2011})}\BibitemShut
  {NoStop}%
\bibitem [{\citenamefont {Boning}(2012)}]{boning2012}%
  \BibitemOpen
  \bibfield  {author} {\bibinfo {author} {\bibfnamefont {S.}~\bibnamefont
  {Boning}},\ }\href
  {http://www.nipne.ro/indico/getFile.py/access?contribId=242&sessionId=3&resId=0&materialId=slides&confId=0}
  {\enquote {\bibinfo {title} {Probing the quadrupole collectivity of 128Cd
  using coulomb excitation},}\ } (\bibinfo {year} {2012}),\ \bibinfo {note}
  {2nd European Nuclear Physics Conference, Bucharest}\BibitemShut {NoStop}%
\bibitem [{\citenamefont {Radford}\ \emph {et~al.}(2005)\citenamefont
  {Radford}, \citenamefont {Baktash}, \citenamefont {Barton}, \citenamefont
  {Batchelder}, \citenamefont {Beene}, \citenamefont {Bingham}, \citenamefont
  {Caprio}, \citenamefont {Danchev}, \citenamefont {Fuentes}, \citenamefont
  {Galindo-Uribarri}, \citenamefont {del Campo}, \citenamefont {Gross},
  \citenamefont {Halbert}, \citenamefont {Hartley}, \citenamefont {Hausladen},
  \citenamefont {Hwang}, \citenamefont {Krolas}, \citenamefont {Larochelle},
  \citenamefont {Liang}, \citenamefont {Mueller}, \citenamefont {Padilla},
  \citenamefont {Pavan}, \citenamefont {Piechaczek}, \citenamefont {Shapira},
  \citenamefont {Stracener}, \citenamefont {Varner}, \citenamefont {Woehr},
  \citenamefont {Yu},\ and\ \citenamefont {Zamfir}}]{radford2005}%
  \BibitemOpen
  \bibfield  {author} {\bibinfo {author} {\bibfnamefont {D.}~\bibnamefont
  {Radford}}, \bibinfo {author} {\bibfnamefont {C.}~\bibnamefont {Baktash}},
  \bibinfo {author} {\bibfnamefont {C.}~\bibnamefont {Barton}}, \bibinfo
  {author} {\bibfnamefont {J.}~\bibnamefont {Batchelder}}, \bibinfo {author}
  {\bibfnamefont {J.}~\bibnamefont {Beene}}, \bibinfo {author} {\bibfnamefont
  {C.}~\bibnamefont {Bingham}}, \bibinfo {author} {\bibfnamefont
  {M.}~\bibnamefont {Caprio}}, \bibinfo {author} {\bibfnamefont
  {M.}~\bibnamefont {Danchev}}, \bibinfo {author} {\bibfnamefont
  {B.}~\bibnamefont {Fuentes}}, \bibinfo {author} {\bibfnamefont
  {A.}~\bibnamefont {Galindo-Uribarri}}, \bibinfo {author} {\bibfnamefont
  {J.~G.}\ \bibnamefont {del Campo}}, \bibinfo {author} {\bibfnamefont
  {C.}~\bibnamefont {Gross}}, \bibinfo {author} {\bibfnamefont
  {M.}~\bibnamefont {Halbert}}, \bibinfo {author} {\bibfnamefont
  {D.}~\bibnamefont {Hartley}}, \bibinfo {author} {\bibfnamefont
  {P.}~\bibnamefont {Hausladen}}, \bibinfo {author} {\bibfnamefont
  {J.}~\bibnamefont {Hwang}}, \bibinfo {author} {\bibfnamefont
  {W.}~\bibnamefont {Krolas}}, \bibinfo {author} {\bibfnamefont
  {Y.}~\bibnamefont {Larochelle}}, \bibinfo {author} {\bibfnamefont
  {J.}~\bibnamefont {Liang}}, \bibinfo {author} {\bibfnamefont
  {P.}~\bibnamefont {Mueller}}, \bibinfo {author} {\bibfnamefont
  {E.}~\bibnamefont {Padilla}}, \bibinfo {author} {\bibfnamefont
  {J.}~\bibnamefont {Pavan}}, \bibinfo {author} {\bibfnamefont
  {A.}~\bibnamefont {Piechaczek}}, \bibinfo {author} {\bibfnamefont
  {D.}~\bibnamefont {Shapira}}, \bibinfo {author} {\bibfnamefont
  {D.}~\bibnamefont {Stracener}}, \bibinfo {author} {\bibfnamefont
  {R.}~\bibnamefont {Varner}}, \bibinfo {author} {\bibfnamefont
  {A.}~\bibnamefont {Woehr}}, \bibinfo {author} {\bibfnamefont {C.-H.}\
  \bibnamefont {Yu}}, \ and\ \bibinfo {author} {\bibfnamefont {N.}~\bibnamefont
  {Zamfir}},\ }\href@noop {} {\bibfield  {journal} {\bibinfo  {journal}
  {Nuclear Physics A}\ }\textbf {\bibinfo {volume} {752}},\ \bibinfo {pages}
  {264 } (\bibinfo {year} {2005})},\ \bibinfo {note} {proceedings of the 22nd
  International Nuclear Physics Conference (Part 2)}\BibitemShut {NoStop}%
\bibitem [{\citenamefont {Beene}\ \emph {et~al.}(2004)\citenamefont {Beene},
  \citenamefont {Varner}, \citenamefont {Baktash}, \citenamefont
  {Galindo-Uribarri}, \citenamefont {Gross}, \citenamefont {del Campo},
  \citenamefont {Halbert}, \citenamefont {Hausladen}, \citenamefont
  {Larochelle}, \citenamefont {Liang}, \citenamefont {Mas}, \citenamefont
  {Mueller}, \citenamefont {Padilla-Rodal}, \citenamefont {Radford},
  \citenamefont {Shapira}, \citenamefont {Stracener}, \citenamefont
  {Urrego-Blanco},\ and\ \citenamefont {Yu}}]{beene2004}%
  \BibitemOpen
  \bibfield  {author} {\bibinfo {author} {\bibfnamefont {J.}~\bibnamefont
  {Beene}}, \bibinfo {author} {\bibfnamefont {R.}~\bibnamefont {Varner}},
  \bibinfo {author} {\bibfnamefont {C.}~\bibnamefont {Baktash}}, \bibinfo
  {author} {\bibfnamefont {A.}~\bibnamefont {Galindo-Uribarri}}, \bibinfo
  {author} {\bibfnamefont {C.}~\bibnamefont {Gross}}, \bibinfo {author}
  {\bibfnamefont {J.~G.}\ \bibnamefont {del Campo}}, \bibinfo {author}
  {\bibfnamefont {M.}~\bibnamefont {Halbert}}, \bibinfo {author} {\bibfnamefont
  {P.}~\bibnamefont {Hausladen}}, \bibinfo {author} {\bibfnamefont
  {Y.}~\bibnamefont {Larochelle}}, \bibinfo {author} {\bibfnamefont
  {J.}~\bibnamefont {Liang}}, \bibinfo {author} {\bibfnamefont
  {J.}~\bibnamefont {Mas}}, \bibinfo {author} {\bibfnamefont {P.}~\bibnamefont
  {Mueller}}, \bibinfo {author} {\bibfnamefont {E.}~\bibnamefont
  {Padilla-Rodal}}, \bibinfo {author} {\bibfnamefont {D.}~\bibnamefont
  {Radford}}, \bibinfo {author} {\bibfnamefont {D.}~\bibnamefont {Shapira}},
  \bibinfo {author} {\bibfnamefont {D.}~\bibnamefont {Stracener}}, \bibinfo
  {author} {\bibfnamefont {J.-P.}\ \bibnamefont {Urrego-Blanco}}, \ and\
  \bibinfo {author} {\bibfnamefont {C.-H.}\ \bibnamefont {Yu}},\ }\href@noop {}
  {\bibfield  {journal} {\bibinfo  {journal} {Nuclear Physics A}\ }\textbf
  {\bibinfo {volume} {746}},\ \bibinfo {pages} {471 } (\bibinfo {year}
  {2004})},\ \bibinfo {note} {proceedings of the Sixth International Conference
  on Radioactive Nuclear Beams (RNB6)}\BibitemShut {NoStop}%
\bibitem [{\citenamefont {Stone}(2011)}]{stone2011}%
  \BibitemOpen
  \bibfield  {author} {\bibinfo {author} {\bibfnamefont {N.~J.}\ \bibnamefont
  {Stone}},\ }\href@noop {} {\emph {\bibinfo {title} {Table of Nuclear Magnetic
  Dipole and Electric Quadrupole Moments}}},\ \bibinfo {type} {Tech. Rep.}\
  \bibinfo {number} {indc(nds)-0594}\ (\bibinfo  {institution} {IAEA, Nuclear
  Data Services},\ \bibinfo {year} {2011})\BibitemShut {NoStop}%
\bibitem [{\citenamefont {LeBlanc}\ \emph {et~al.}(2005)\citenamefont
  {LeBlanc}, \citenamefont {Cabaret}, \citenamefont {Cottereau}, \citenamefont
  {Crawford}, \citenamefont {Essabaa}, \citenamefont {Genevey}, \citenamefont
  {Horn}, \citenamefont {Huber}, \citenamefont {Lassen}, \citenamefont {Lee},
  \citenamefont {Scornet}, \citenamefont {Lettry}, \citenamefont {Obert},
  \citenamefont {Oms}, \citenamefont {Ouchrif}, \citenamefont {Pinard},
  \citenamefont {Ravn}, \citenamefont {Roussi\`ere}, \citenamefont {Sauvage},\
  and\ \citenamefont {Verney}}]{leblanc2005}%
  \BibitemOpen
  \bibfield  {author} {\bibinfo {author} {\bibfnamefont {F.}~\bibnamefont
  {LeBlanc}}, \bibinfo {author} {\bibfnamefont {L.}~\bibnamefont {Cabaret}},
  \bibinfo {author} {\bibfnamefont {E.}~\bibnamefont {Cottereau}}, \bibinfo
  {author} {\bibfnamefont {J.~E.}\ \bibnamefont {Crawford}}, \bibinfo {author}
  {\bibfnamefont {S.}~\bibnamefont {Essabaa}}, \bibinfo {author} {\bibfnamefont
  {J.}~\bibnamefont {Genevey}}, \bibinfo {author} {\bibfnamefont
  {R.}~\bibnamefont {Horn}}, \bibinfo {author} {\bibfnamefont {G.}~\bibnamefont
  {Huber}}, \bibinfo {author} {\bibfnamefont {J.}~\bibnamefont {Lassen}},
  \bibinfo {author} {\bibfnamefont {J.~K.~P.}\ \bibnamefont {Lee}}, \bibinfo
  {author} {\bibfnamefont {G.~L.}\ \bibnamefont {Scornet}}, \bibinfo {author}
  {\bibfnamefont {J.}~\bibnamefont {Lettry}}, \bibinfo {author} {\bibfnamefont
  {J.}~\bibnamefont {Obert}}, \bibinfo {author} {\bibfnamefont
  {J.}~\bibnamefont {Oms}}, \bibinfo {author} {\bibfnamefont {A.}~\bibnamefont
  {Ouchrif}}, \bibinfo {author} {\bibfnamefont {J.}~\bibnamefont {Pinard}},
  \bibinfo {author} {\bibfnamefont {H.}~\bibnamefont {Ravn}}, \bibinfo {author}
  {\bibfnamefont {B.}~\bibnamefont {Roussi\`ere}}, \bibinfo {author}
  {\bibfnamefont {J.}~\bibnamefont {Sauvage}}, \ and\ \bibinfo {author}
  {\bibfnamefont {D.}~\bibnamefont {Verney}},\ }\href@noop {} {\bibfield
  {journal} {\bibinfo  {journal} {Phys. Rev. C}\ }\textbf {\bibinfo {volume}
  {72}},\ \bibinfo {pages} {034305} (\bibinfo {year} {2005})}\BibitemShut
  {NoStop}%
\bibitem [{\citenamefont {Bohr}\ and\ \citenamefont
  {Mottelson}(1975)}]{bm-vol2-1975}%
  \BibitemOpen
  \bibfield  {author} {\bibinfo {author} {\bibfnamefont {A.}~\bibnamefont
  {Bohr}}\ and\ \bibinfo {author} {\bibfnamefont {B.~R.}\ \bibnamefont
  {Mottelson}},\ }\href@noop {} {\emph {\bibinfo {title} {Nuclear Structure,
  Volume II: Nuclear Deformations}}}\ (\bibinfo  {publisher} {W.A. Benjamin,
  Inc},\ \bibinfo {year} {1975})\BibitemShut {NoStop}%
\end{thebibliography}
%

\end{document}